\documentclass[twocolumn,showpacs,preprintnumbers,amsmath,amssymb,prl]{revtex4}

\usepackage{graphicx}
\usepackage{amssymb}
\usepackage{dcolumn}
\usepackage{bm}
\usepackage{epstopdf}
\usepackage{wasysym}
\usepackage{ulem}
\usepackage{MnSymbol}
\usepackage{color}
\newcommand{\tildeleft}{{\raise.17ex\hbox{$\scriptstyle\sim$}}}

\begin{document}
\title{Roton-Maxon Excitation Spectrum of Bose Condensates in a Shaken Optical Lattice}
\author{Li-Chung Ha$^1$, Logan W. Clark$^1$, Colin V. Parker$^1$, Brandon M. Anderson$^{1,2}$, Cheng Chin$^1$}
\affiliation{$^1$James Franck Institute, Enrico Fermi Institute and Department of Physics, University of Chicago, Chicago, IL 60637, USA}
\affiliation{$^2$Joint Quantum Institute, University of Maryland, College Park, MD 20742, USA}
\date{\today}

\begin{abstract}
We present experimental evidence showing that an interacting Bose condensate in a shaken optical lattice develops a roton-maxon excitation spectrum, a feature normally associated with superfluid helium. The roton-maxon feature originates from the double-well dispersion in the shaken lattice, and can be controlled by both the atomic interaction and the lattice modulation amplitude. We determine the excitation spectrum using Bragg spectroscopy and measure the critical velocity by dragging a weak speckle potential through the condensate -- both techniques are based on a digital micromirror device. Our dispersion measurements are in good agreement with a modified Bogoliubov model.
\end{abstract}


\maketitle

In his seminal papers in the 1940s \cite{Landau41,Landau47}, L.~D. Landau formulated the theory of superfluid helium-4 (He II) and showed that the energy-momentum relation (dispersion) of He II supports two types of elementary excitations: acoustic phonons and gapped rotons. This dispersion underpins our understanding of superfluidity in helium, and explains many experiments on heat capacity and superfluid critical velocity. What is now called the ``roton-maxon'' dispersion in He II has been precisely measured in neutron scattering experiments \cite{Henshaw61, Glyde94} and is generally considered a hallmark of Bose superfluids in the strong interaction regime.

The roton-maxon dispersion carries a number of intriguing features that distinguish excitations in different regimes. The low-lying excitations are acoustic phonons with energy $E=p v_s$, where $p$ is the momentum and $v_s$ is the sound speed. At higher momenta, the dispersion exhibits both a local maximum at $p=p_m$ with energy $E=\Delta_m$ and a minimum at $p=p_r$ with energy $E=\Delta_r$. The elementary excitations associated with this maximum and minimum are known as maxons and rotons, respectively. The roton excitations, in particular, are known to reduce the superfluid critical velocity below the sound speed. This is best understood based on the Landau criterion for superfluidity in which the critical velocity set by the roton minimum $v_c\approx \Delta_r/p_r$ is lower than the sound speed $v_s$. The roton minimum also suggests the emergence of density wave order \cite{Schneider71} and dynamical instability \cite{Santos2003}.

To explore the properties of these unconventional excitations, many theoretical works have proposed schemes for producing the roton-maxon dispersion outside of the He II system. Many proposals have been devoted to atomic systems with long-range or enhanced interactions, e.g. dipolar gases \cite{Santos2003, Wilson2008, ODell2003}, Rydberg-excited condensates \cite{Rb_roton}, or resonantly interacting gases \cite{Fb_roton}. Other candidates are 2D Bose gases \cite{Fischer06, Nogueira06}, spinor condensates \cite{Cherng09, Matuszewski10}, and spin-orbit coupled condensates \cite{Higbie02,SOC}. Experimentally, mode softening resulting from cavity-induced interaction has recently been reported \cite{Mottl12}, which provides strong evidence for an underlying rotonlike excitation spectrum.

In this Letter, we generate and characterize an asymmetric roton-maxon excitation spectrum based on a Bose-Einstein condensate (BEC) in a one dimensional (1D) shaken optical lattice. We implement Bragg spectroscopy and identify the local maximum and minimum in the dispersion associated with the maxon and roton excitations. Furthermore, by dragging a speckle potential through the BEC we show a reduction of the superfluid critical velocity in the presence of the roton dispersion.
 
We create the roton-maxon dispersion by loading a 3D Bose condensate into a 1D shaken (i.e. periodically phase modulated) optical lattice. The lattice shaking technique has been used previously to engineer novel band structures \cite{Gemelke05,Lignier07} and to simulate magnetism \cite{Struck11,Parker13,Jotzu14}. Here, we phase modulate the lattice to create a double-well structure in the single-particle dispersion $\epsilon_0(q)$, for which the ground state has a twofold degeneracy; see Fig.~\ref{fig1}(a) and Ref.~\cite{Parker13}. The double-well dispersion results from a near resonant coupling between the ground and first excited band through lattice shaking \cite{Parker13}, and is a consequence of the parametric instability of a driven anharmonic oscillator \cite{Gemelke05}. The dispersion with quasimomentum $q$ can be calculated based on a Floquet model \cite{Parker13}. A similar double-well dispersion can also be realized in a spin-orbit coupled system \cite{Lin11,Cheuk12,Hamner14,Ji14,Fu14}.

\begin{figure}[t]
\hspace*{-0.7cm}\includegraphics[width=2.8 in]{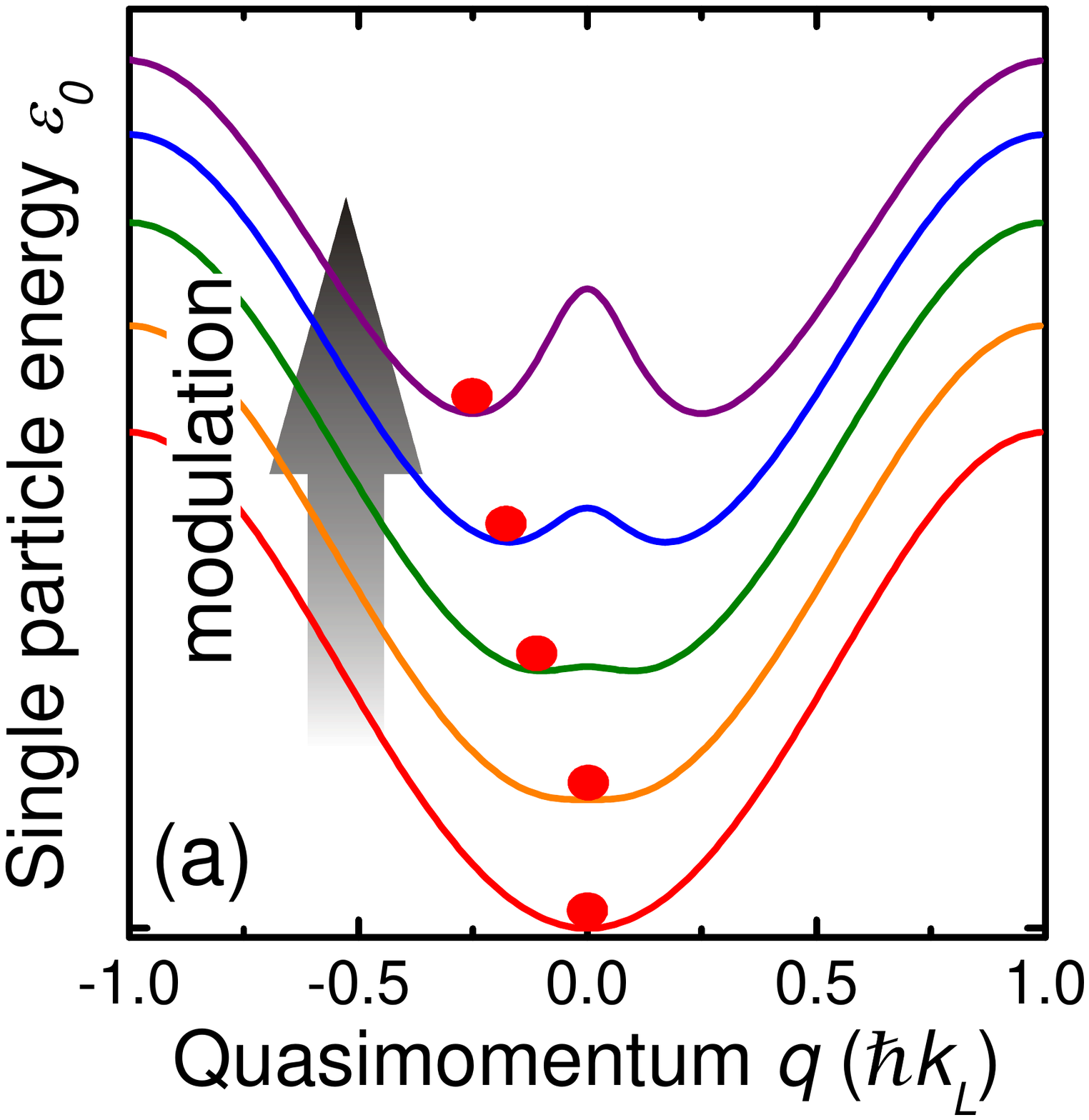}\hspace*{-2.9cm}
\includegraphics[width=2.8 in]{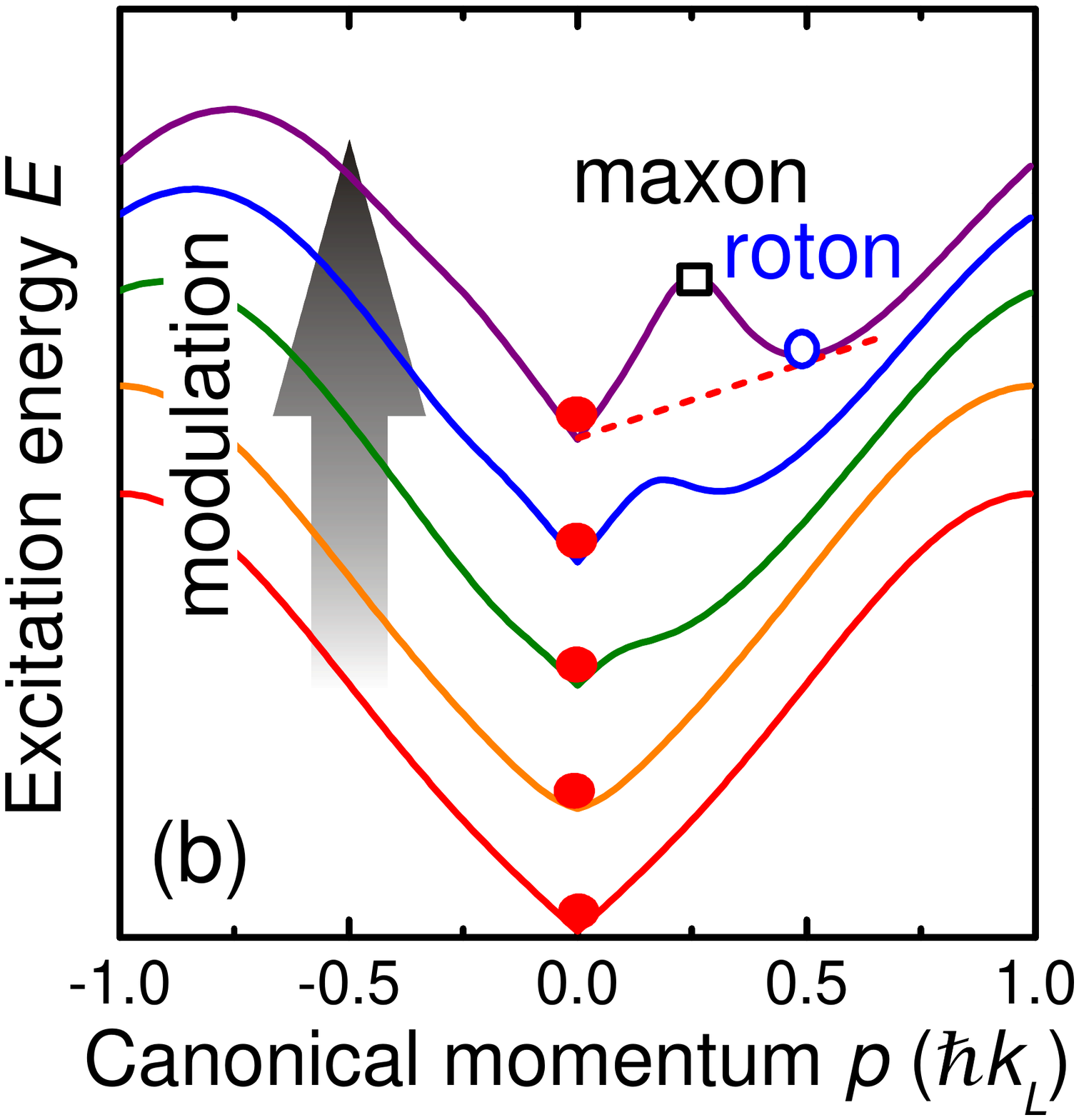}
\caption{{(color online) Generation of roton-maxon dispersion in a shaken lattice.} (a) For a single atom, the lattice modulation creates a double-well structure above a critical modulation amplitude (top three lines) \cite{Parker13}. In our experiment, the atoms are prepared at the minimum with zero or negative momentum ($q^*\leq 0$, red dot); see text. (b) With atomic interactions, a roton minimum (circle) and a maxon maximum (square) in the excitation spectrum can form. The dashed line indicates the critical velocity limited by the roton minimum according to the Landau criterion for superfluidity. Dispersions are upward offset with increasing modulation amplitude for clarity. The lattice reciprocal momentum is $\hbar k_L=h/\lambda$ where $\lambda$ is the wavelength of the lattice beams and $h=2\pi\hbar$ is the Planck constant.}\label{fig1}
\end{figure}

The double-well dispersion is modified by atomic interactions. Assuming the BEC is loaded into one of the two dispersion minima at quasimomentum $q=q^*$, we introduce the canonical momentum $p=q-q^*$ in the reference frame where the condensate has zero momentum and energy. The new dispersion is $\tilde{\epsilon}_0(p)=\epsilon_0(p+q^*)-\epsilon_0(q^*)$. One finds that the dispersion are no longer symmetric due to the existence of the other unoccupied minimum; see Fig.~\ref{fig1}(b). Based on a modified Bogoliubov calculation (see  Supplemental Material \cite{Supp} and Refs.~\cite{Eckardt10, Struck13}), we diagonalize the Hamiltonian to obtain the excitation spectrum:

\begin{eqnarray}\label{eq2}
E(p)=\sqrt{\bar{\epsilon}(p)^2+2\mu\bar{\epsilon}(p)}+\Delta \epsilon(p),
\end{eqnarray}
where $\bar{\epsilon}(p)=[\tilde{\epsilon}_0(p)+\tilde{\epsilon}_0(-p)]/2$, $\Delta \epsilon(p)=[\tilde{\epsilon}_0(p)-\tilde{\epsilon}_0(-p)]/2$ and $\mu$ is the chemical potential. For a system with a double-well structure in $\tilde{\epsilon}_0(p)$, the theory predicts a roton-maxon structure with the roton minimum occuring near $p=-2q^*$; see Fig.~\ref{fig1}(b). Creation of an \textgravedbl artificial roton\textacutedbl ~in the dispersion minimum of an analogous spin-orbit coupled system was theoretically proposed in Ref.~\cite{Higbie02}.

\begin{figure}[t]
\includegraphics[width=3.5 in]{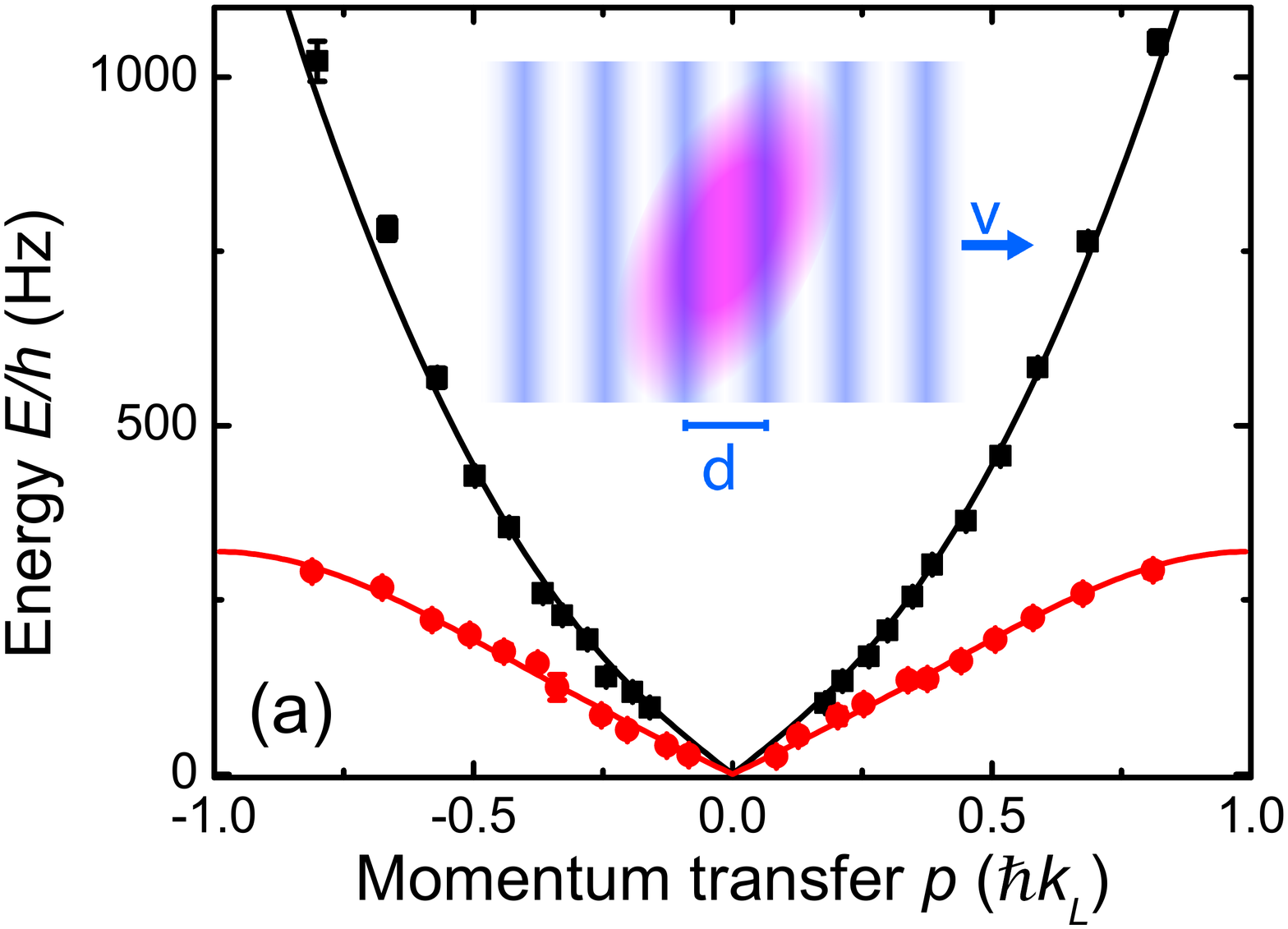}\\ \vspace*{-0.4cm}
\includegraphics[width=3.5 in]{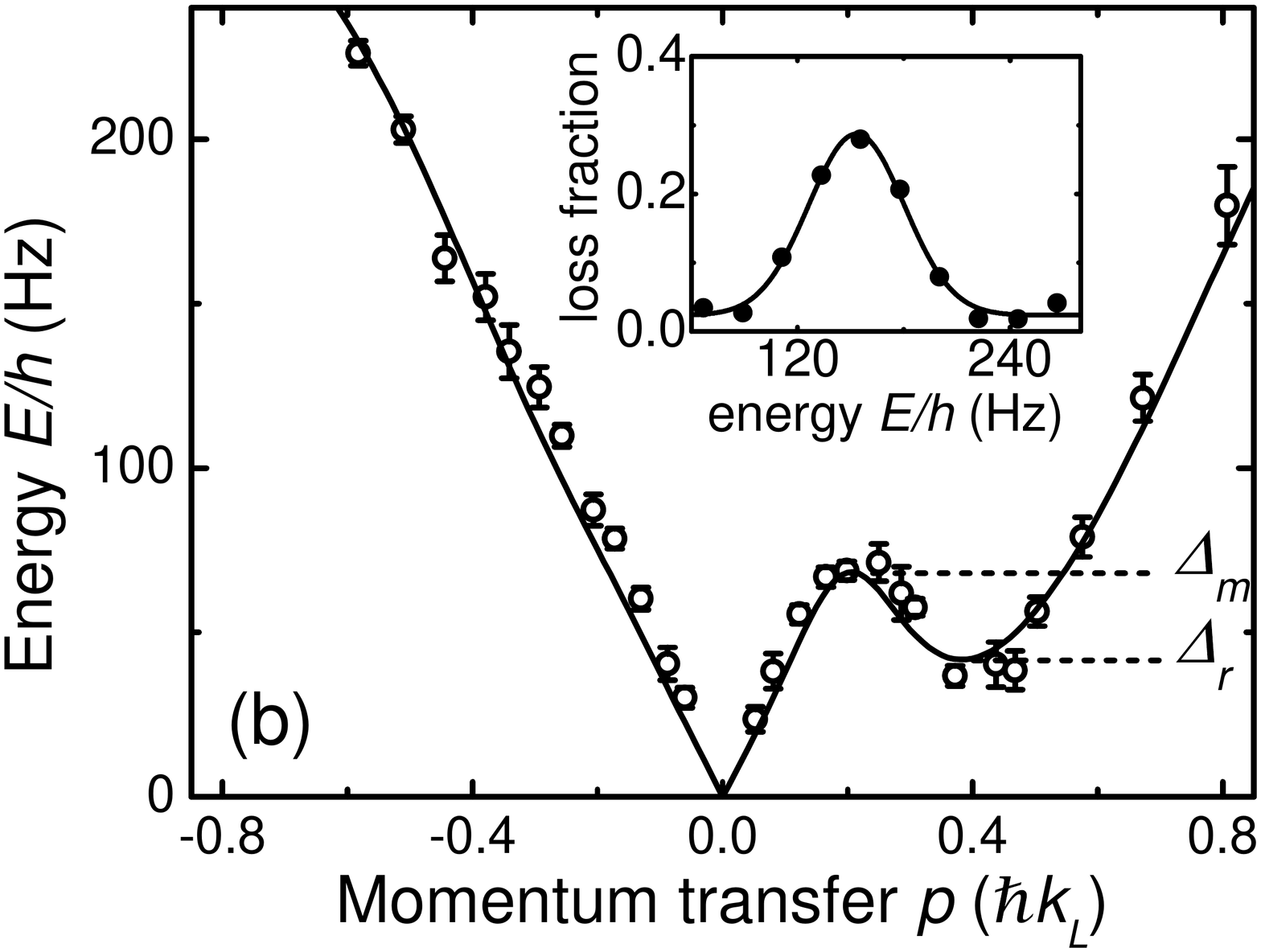}
\caption{{(color online) Excitation spectra.} (a) We measure the excitation spectra with $N_0=30~000$ atoms in a harmonic trap (square) and in a stationary lattice (circle) with DMD-based Bragg spectroscopy. The inset illustrates the moving optical potential with velocity $v$ and periodicity $d$ created by the DMD on the BEC (tilted ellipse); see text. The solid lines correspond to the Bogoliubov model with chemical potentials equal to the trap-averaged values. (b) For a BEC with $N_0=9000$ atoms loaded in a shaken optical lattice, we measure the excitation spectrum along the lattice direction. The modulation amplitude (peak to peak) is $\Delta x= 33~$nm. The solid line is the best fit based on Eq.~(1). The inset shows a typical atom loss spectrum taken at $k=-0.38~k_L$. In both panels, the scattering length is $a=47~a_0$.}\label{fig2}
\end{figure}

Our experiment to detect this unusual dispersion starts with an almost pure cesium condensate of $N_0=30~000$ atoms loaded into a  crossed beam optical dipole trap (wavelength $\lambda=1064$~nm) with trap frequencies $(\omega_x, \omega_y, \omega_z)=2\pi \times (9.3, 27, 104)~$Hz \cite{Parker13}. We turn on an additional 1D optical lattice by retroreflecting one of the dipole trap beams in the $x-y$ plane at $40^\circ$ with respect to the $x$-axis. The lattice depth is approximately $V=7~E_R$, where $E_R=h\times 1.325~$kHz is the photon recoil energy of the lattice beam. The lattice potential is phase modulated at 7.3~kHz which is 0.7~kHz blue detuned from the ground to first excited band transition at $q=0$. The phase modulation creates admixed bands, and the ground band develops two minima in its dispersion \cite{Parker13}. We preferentially load the BEC into one of the minima by providing a momentum kick before phase modulating the lattice \cite{Parker13}. We define the direction of the kick as negative, and thus the BEC has a negative momentum $q=q^*<0$ and the roton minimum is expected at $p=2|q^*|$; see Fig.~\ref{fig1}(b).

To probe the dispersion we perform Bragg spectroscopy \cite{Ozeri05} by illuminating the atoms with a sinusoidal potential moving along the direction of the shaken lattice. The potential is created from a programmable digital micromirror device (DMD) and a 789~nm laser, which provides a repulsive dipole force. The DMD potential with velocity $v$ and periodicity $d$ [see Fig.~\ref{fig2}(a) inset] induces a Raman coupling between the condensate with $p=0$ and finite momentum states with $p=h/d$. When the Raman detuning $E=pv$ matches the energy of the finite momentum state $E(p)$, a resonant transfer will remove atoms from the condensate. We illuminate the atoms with the moving potential for 40~ms and measure the residual condensate particle number after a 30~ms time of flight (TOF). The dispersion can be mapped out by finding the energy which gives the strongest reduction of atom number in the condensate for each momentum $p$.

To test this technique, we compare the dispersions of the BEC in a harmonic trap and that in a $V=7~E_R$ unshaken lattice to Bogoliubov calculations; see Fig.~\ref{fig2}(a). The measurement agrees well with the Bogoliubov spectrum using the measured trap-averaged chemical potentials $\mu= h\times 120~$Hz without the lattice and $\mu= h\times 150~$Hz with the lattice.

We now consider the dispersion of a BEC in a shaken optical lattice, where the roton feature is expected. Here we observe a distinct difference between the excitations at positive versus negative momentum. We work with a modulation amplitude (peak to peak) of $\Delta x= 33$~nm which guarantees a strong double-well feature. Fig.~\ref{fig2}(b) shows the dispersion measurement, which contains a clear roton-maxon feature at positive momentum (hereafter, the roton direction). In contrast, we do not see this feature for negative momentum (hereafter the nonroton direction).

We compare the measured roton spectrum with the model in Eq.~(1). Constraining the model to the experimental parameters only yields qualitative agreement likely due to interaction effects \cite{Wei14} which effectively modify the modulation amplitude $\Delta x$ and lattice depth $V$. Thus we fit the data with Eq.(1) and find the best fit to have $\mu= h\times 58(4)~$Hz, $V=5.9(1)~E_R$ and $\Delta x = 49(3)~$nm. The low chemical potential is expected and comes from the lower condensate number as well as the weaker, momentum dependent atomic interactions in the admixed band.

\begin{figure}[t]
\includegraphics[width=3.5 in]{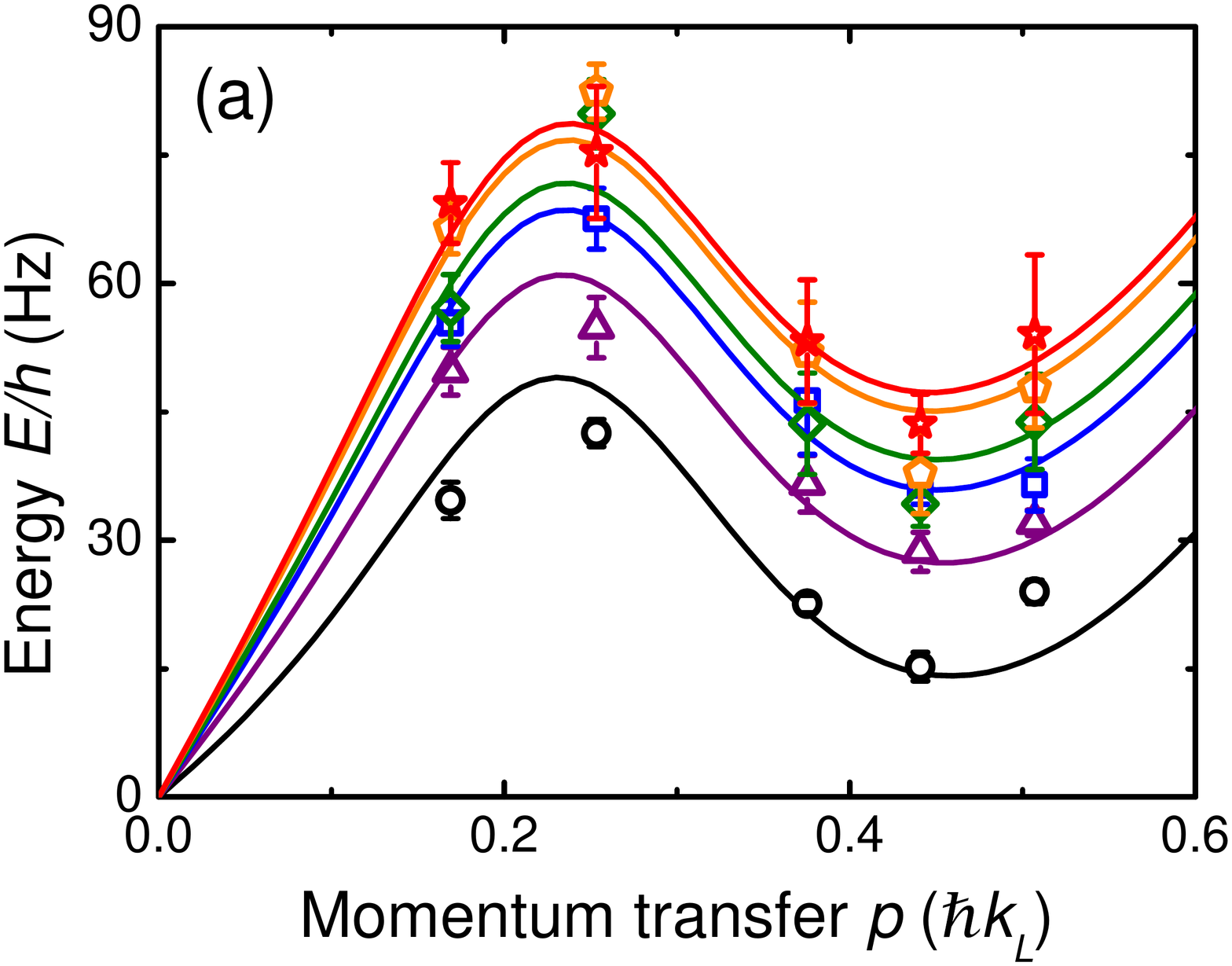}\\ \vspace*{-0.4cm}
\includegraphics[width=3.5 in]{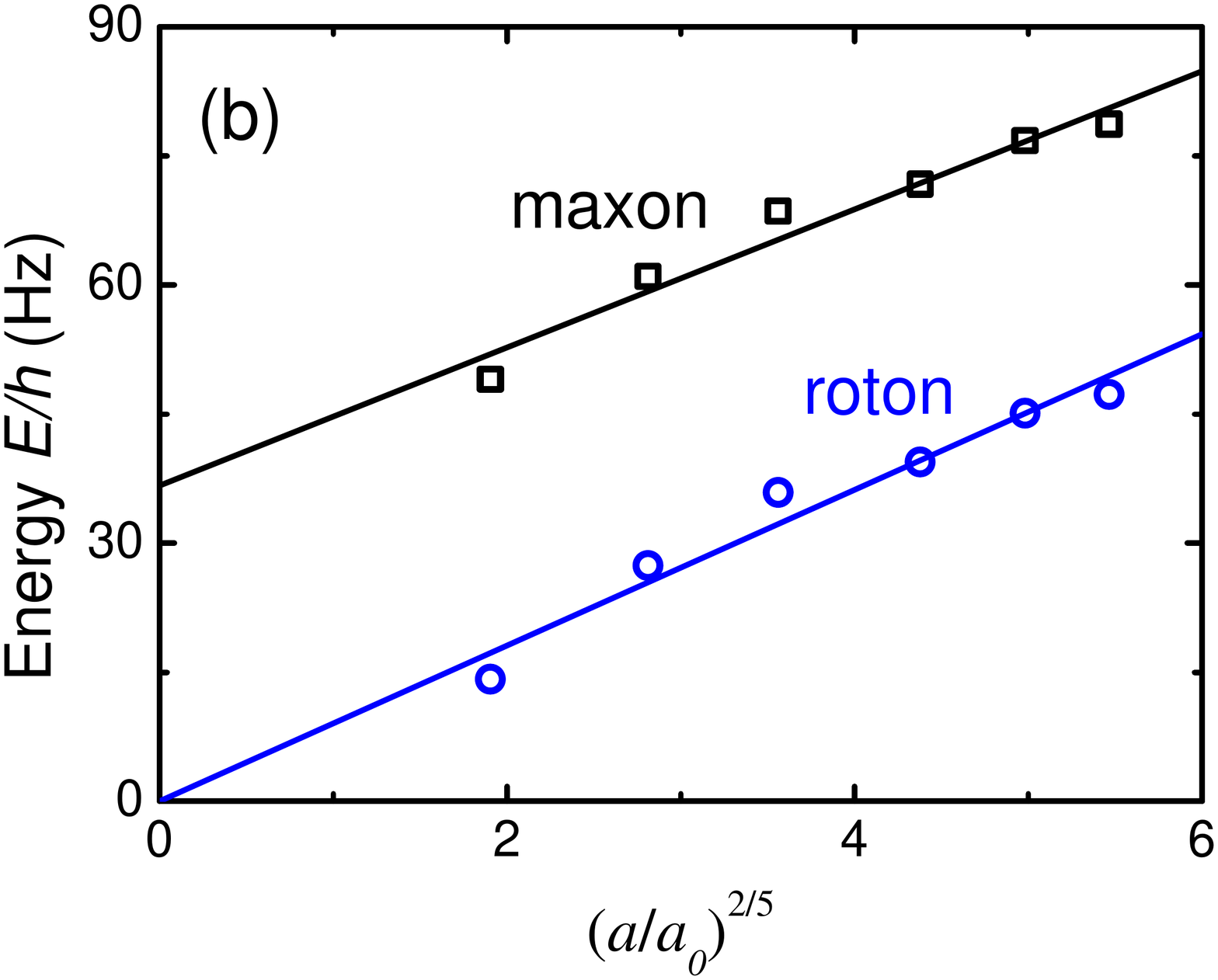}
\caption{{(color online) Roton or maxon energy vs scattering length.} (a) We measure the excitation spectra at different scattering lengths $a/a_0$=5 (circle), 13 (triangle), 24 (square), 40 (diamond), 55 (pentagon) and 70 (star). The condensate number is $N_0=9000$. Solid curves are fits based on Eq.~(1). A global optimization procedure gives $V=6.7(2)~E_R$ and $\Delta x=43(3)$~nm. (b) Roton energies (circle) and maxon energies (square) extracted from the fits in panel (a) are shown at different scattering lengths. Solid curves are fits based on $\Delta_r=A (a/a_0)^{2/5}$ and $\Delta_m=B+C (a/a_0)^{2/5}$, from which we obtain $A=h\times 9(1)~$Hz, $B=h\times 37(9)~$Hz and $C=h\times 8(1)~$Hz.}\label{fig3}
\end{figure}

The roton energy is determined by atomic interactions and can be controlled by tuning the scattering length. To demonstrate this we prepare samples with the usual procedure but at a higher scattering length $a=70~a_0$ followed by ramping the magnetic field to reach the desired scattering length \cite{Chin10}. We measure the excitation spectrum in the roton direction with $p>0$ at six different scattering lengths, shown in Fig.~\ref{fig3}(a).

We adopt a global fit to the data in Fig.~\ref{fig3}(a) based on Eq.~(1) to determine the roton energy $\Delta_r$ and the maxon energy $\Delta_m$. Our observation shows that we can experimentally tune the scattering length to vary the roton energy by a factor of 3. Furthermore, we can use scaling arguments to distinguish the behavior of rotons and maxons from the more conventional phonons. For small chemical potentials, the excitation energy for phonons is well known to scale as $\mu^{1/2}$, while the roton and maxon energies are expected to depend linearly on $\mu$; see Supplemental Material \cite{Supp}. Furthermore, for an adiabatic ramp of the scattering length, the chemical potential should scale as $\mu=n_0g \propto a^{2/5}$ where $g\propto a$ is the interaction strength, and the condensate density in the harmonic trap is $n_0\propto a^{-3/5}$ \cite{Pethick}. Therefore, we plot the extracted roton and maxon energies as a function of $a^{2/5}$ as a proxy for the chemical potential; see Fig.~\ref{fig3}(b). The observed linear dependence confirms the expected scaling for rotons and maxons.

One significant consequence of the roton dispersion is the suppressed superfluid critical velocity $v_c$. We measure the critical velocity of the BEC loaded into the shaken lattice by projecting a moving speckle pattern using the DMD. Instead of using a single laser beam \cite{Raman99,Onofrio00,Desbuquois12} or a lattice with a definite spatial frequency \cite{Miller07}, our speckle pattern contains a broad spectrum of wavenumbers up to the resolution ($k\approx 0.55~k_L$) of our projection system. Furthermore, the potential remains locally perturbative ($\approx h\times 1.1$~Hz) to prevent vortex proliferation \cite{Frisch92, Stiesberger00, Winiecki00}. When the velocity of the speckle pattern reaches or exceeds the critical velocity, atoms are excited from the condensate. To prevent excitation in the low density tail \cite{Miller07}, we digitally mask out the region of the speckle pattern which could overlap with the edge of the cloud.

We observe a clear threshold in speckle velocity above which the condensate number decreases; see Fig.~\ref{fig4}(a). The experimental sequence is similar to that used for Bragg spectroscopy: we illuminate the cloud with a moving speckle pattern for 100~ms followed by a 30~ms TOF. To find the critical velocity, we fit the remaining condensate number with a constant value intersecting a linear decay. The intersection point determines the critical velocity $v_c$. Above a critical value, we observe the condensate fraction decreases linearly with the speckle velocity. This is consistent with a previous observation of the critical velocity in a Bose superfluid \cite{Miller07}, along with a recent calculation \cite{Baym12}.

\begin{figure}[t]
\includegraphics[width=3.5 in]{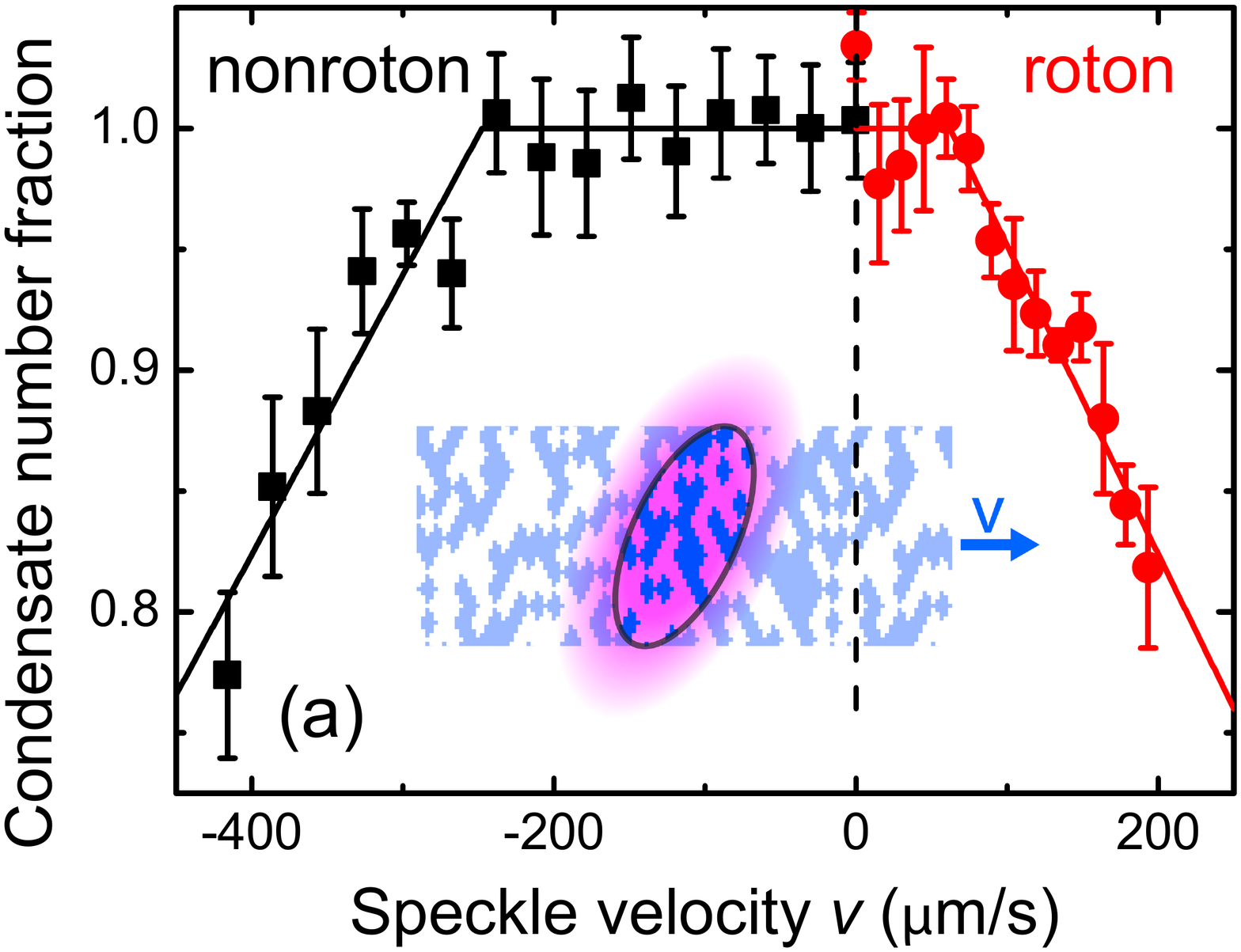}\\ \vspace*{-0.4cm}
\includegraphics[width=3.5 in]{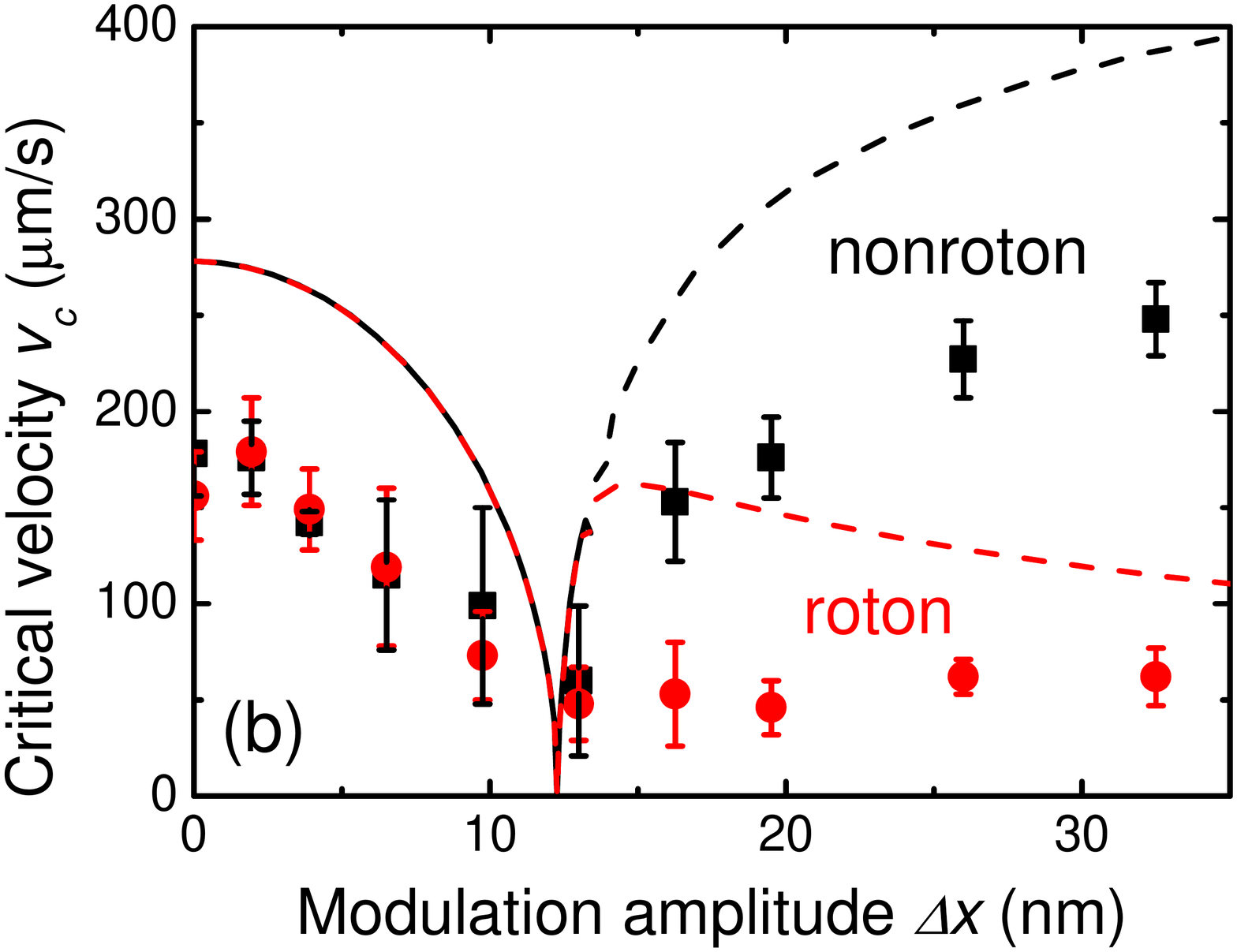}
\caption{{(color online) Superfluid critical velocity.} (a) We measure the residual condensate number fraction after dragging a speckle pattern through the center of the cloud at different velocities $v$ along the roton direction ($p>0$, solid dots) and the nonroton direction ($p<0$, solid squares). The solid lines are fits used to determine the critical velocity. The inset illustrates the experimental scheme; see text. (b) Critical velocities as a function of modulation amplitude are shown. Above the critical modulation amplitude $\Delta x> 12$~nm, the critical velocity is significantly lower in the roton direction. Our measurement is compared with the critical velocity calculated from Eq.~(1) using Landau criterion (dashed lines). In both panels, the scattering length is $a=47~a_0$ and the initial condensate number is $N_0=9000$.}\label{fig4}
\end{figure}

In order to understand the emergence of the roton-maxon dispersion, we measure critical velocity in both the roton direction $p>0$ and the nonroton direction $p<0$ with increasing modulation amplitude $\Delta x$; see Fig.~\ref{fig4}(b). In order to maintain a comparable chemical potential, we prepare the samples with a large $\Delta x = 33~$nm and slowly ramp $\Delta x$ to the desired value. For small final $\Delta x< 12$~nm, $v_c$ is the same in both directions and decreases as we approach the critical value $\Delta x_c\approx 12~$nm (phonon mode softening). When the gas enters the ferromagnetic phase ($\Delta x > 12~$nm) \cite{Parker13}, $v_c$ increases immediately along the nonroton direction, while in the roton direction $v_c$ remains small. 

We compare the measurement with the critical velocity based on the Landau criterion $v_L=\mathrm{min}|E(p)/p|$. As the experiment conditions closely resemble those in Fig.~\ref{fig2}(b), we evaluate the critical velocity with $\mu= h\times 58~$Hz, $V=5.9~E_R$ and $\Delta x$ scaled by $1.5$, the parameters which best fit that dispersion measurement. The calculated $v_L$, shown as dashed lines in Fig.~\ref{fig4}(b), displays a disparity between the roton and nonroton directions for $\Delta x > 15$~nm, in agreement with our observation. Our critical velocities, however, are significantly lower than $v_L$. In early BEC experiments \cite{Raman99,Onofrio00}, low critical velocities were observed and explained by the large obstacles that disrupt the superflow and spin off vortices \cite{Frisch92, Stiesberger00, Winiecki00}. In our experiment with weak speckle potential, a likely scenario is that the critical velocity is limited by excitations generated in the low density regions above and below the cloud along the DMD projection axis.

In conclusion, we observe a roton-maxon dispersion of a BEC in a shaken 1D optical lattice based on three pieces of evidence: the many-body excitation spectrum, the dependence of the excitation energies on the atomic interactions, and the superfluid critical velocity measurement. Our results agree well with the Bogoliubov calculation and suggest the roton/maxon excitations are distinct from acoustic phonons. Our experiment demonstrates that shaken optical lattices are a convenient platform to generate new types of quasiparticles in a dilute atomic gas, allowing future study of their dynamics, stability, and interactions. For instance, knowing the quasiparticle dispersion should allow a future experiment to create macroscopic numbers of rotons, leading to possible roton condensation \cite{Baym12, Pitaevskii84}, and separation of the rotons into domains. \textit{In situ} imaging would allow direct observation of the temporal evolution of such states.

We thank K. Jim\'{e}nez Garc\'{i}a for discussion and careful reading of the article, and U. Eismann and E. Hazlett for assistance in the early phase of the experiment. L.-C. H. is supported by the Grainger Fellowship and the Taiwan
Government Scholarship. L. W. C. is supported by the NDSEG Fellowship. This work was supported by NSF MRSEC Grant No. DMR-1420709, NSF Grant No. PHY-0747907 and ARO-MURI Grant No. W911NF-14-1-0003.

\setcounter{equation}{0}
\setcounter{figure}{0}
\setcounter{table}{0}
\makeatletter
\renewcommand{\theequation}{S\arabic{equation}}
\renewcommand{\thefigure}{S\arabic{figure}}
\renewcommand{\bibnumfmt}[1]{[S#1]}
\renewcommand{\citenumfont}[1]{S#1}

\newpage

\maketitle
\section{Supplemental Material}

\subsection{Optical Setup and Digital Micromirror Device (DMD)}
\begin{figure}[tbh]
\includegraphics[width=3.2 in]{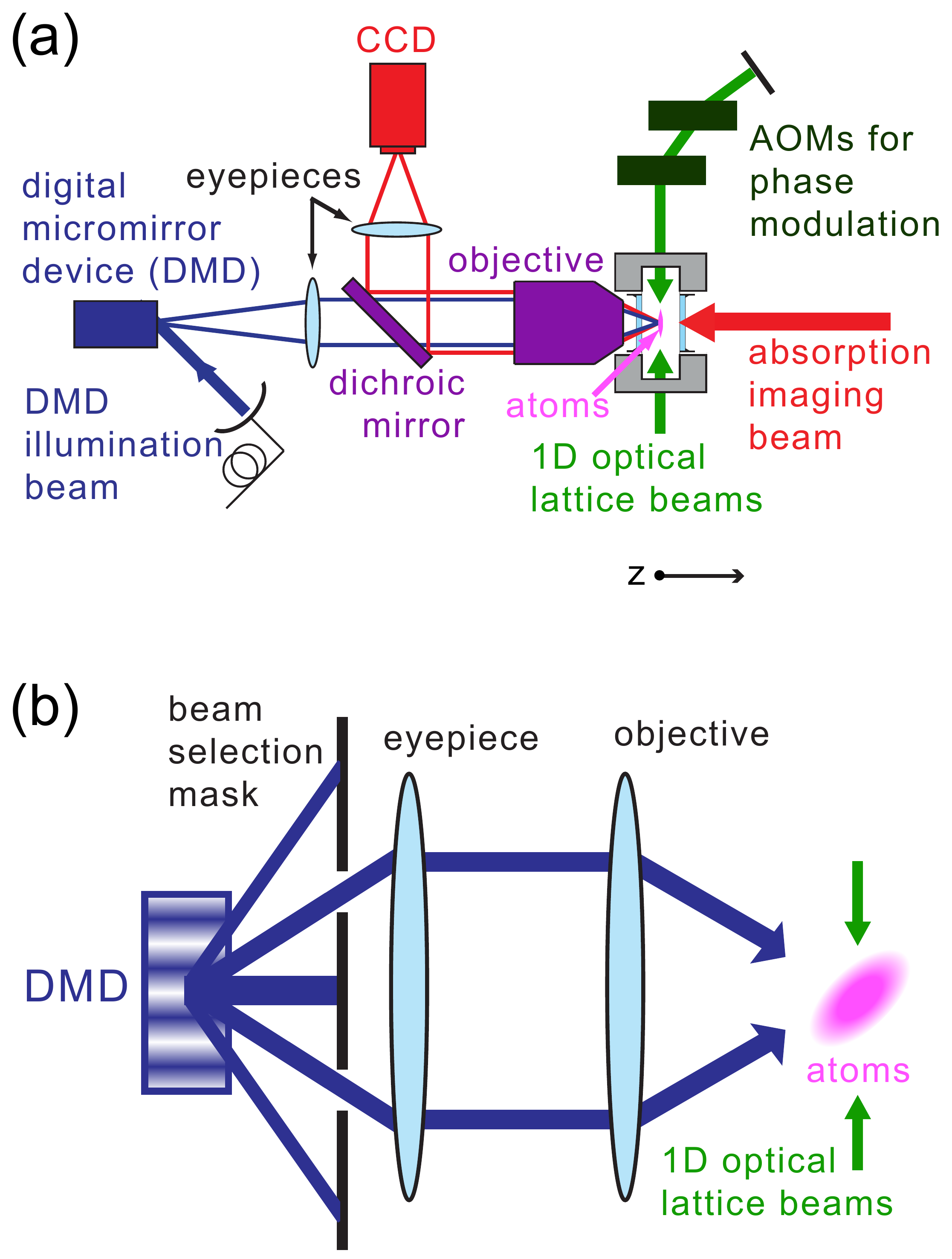}\\
\caption{\textbf{Optical setup} (a) Our optical setup is based on a single high resolution objective lens, which allows us both to perform absorption imaging and to project arbitrary patterns from the DMD onto the atoms (see supplemental text for details). The 1D optical lattice is formed by retroreflecting one of the 1064~nm dipole trap beams after passing it through two acousto-optic modulators (AOMs), which can be used to phase modulate the lattice \cite{Parker13S}. (b) In order to project clean optical potentials for probing dispersion, beam selection masks transmit only the two desired beams diffracted from the DMD. In this way the interference pattern on the atoms contains no confounding wavevectors. \label{figS1}}
\end{figure}

The optical setup that we use for creating and probing the roton-maxon dispersion is shown in Fig.~\ref{figS1}(a). The 1D optical lattice is created by retroreflecting one of the 1064 nm dipole trap beams and is phase modulated using a pair of acousto-optic modulators, as described in our previous work \cite{Parker13S}. We phase modulate the lattice at 7.3~kHz, 0.7 kHz blue detuned from the ground to first excited band transition at $q=0$. This modulation couples the two bands and produces a roton-maxon dispersion.

We have implemented a digital micromirror device (DMD: Texas Instruments, DLP LightCrafter 3000) to tailor dynamic optical potentials for probing the dispersion. The DMD consists of a $608\times 684$ array of $7.6~\mu$m square mirrors. Each mirror flips individually to one of two angles, separated by $24^\circ$. A mirror at the ``on'' angle will reflect light towards the atom cloud, while the ``off'' angle reflects light into a beam dump. We reflect a blue-detuned 789~nm laser off of the DMD and use a high-resolution objective lens to project the real space pattern of mirrors in the ``on'' state onto the plane of the atom cloud. We use additional lenses to demagnify the pattern by a factor of 36. The resolution of the resulting patterns is limited by the objective lens to $\tildeleft 1~\mu$m, approximately 5 micromirrors across. By having many micromirrors in each resolution sized area we can generate intermediate intensities in static patterns even though the state of each micromirror is binary. The programmed pattern of ``on'' mirrors can be updated up to 4000 times per second, allowing us to create motion in the projected patterns.

\subsection{Bragg Spectroscopy with the DMD}
In order to create single-wavevector moving lattices for Bragg spectroscopy, we add additional beam selection masks in front of the DMD as shown in Fig.~\ref{figS1}(b). An approximately sinusoidal DMD pattern will diffract the laser into the desired beams as well as many harmonics. By always transmitting only two diffracted beams through a mask we ensure that the projected optical potential contains only the desired wavevector for probing the atoms. Shifting the pattern of the ``on'' mirrors translates the projected potential by the same amount regardless of the blocking mask. The second important advantage of beam selection is that removing the 0th order diffraction allows us to double the maximum projected wavevector. The 0th order diffracted beam is always more intense than the $\pm1$st, so when all are present the projected potential is dominated by the interference wavevector between the 0th and $\pm1$st. That wavevector is limited by the objective lens to approximately $k=0.55~k_{\rm{L}}$. Once the 0th order beam is blocked, the interference between the 1st and -1st dominates, which raises the maximum projected wavevector to approximately $k=1.1~k_{\rm{L}}$ and is sufficient for our experiments.

For any wavevector of the projected potential, the quasiparticle excitation frequency corresponds to the rate at which the pattern's phase shifts by $2\pi$. We typically use sets of 9 lattice patterns, so that each pattern switch corresponds to a phase change of $2\pi/9$. We scan the excitation frequency by changing the rate at which we trigger the DMD to cycle through the set of patterns. To make the movement smoother for probing the small excitation frequencies in Fig.~3, we use sets of 20 patterns instead.

\begin{figure}[tbh]
\includegraphics[width=3.2 in]{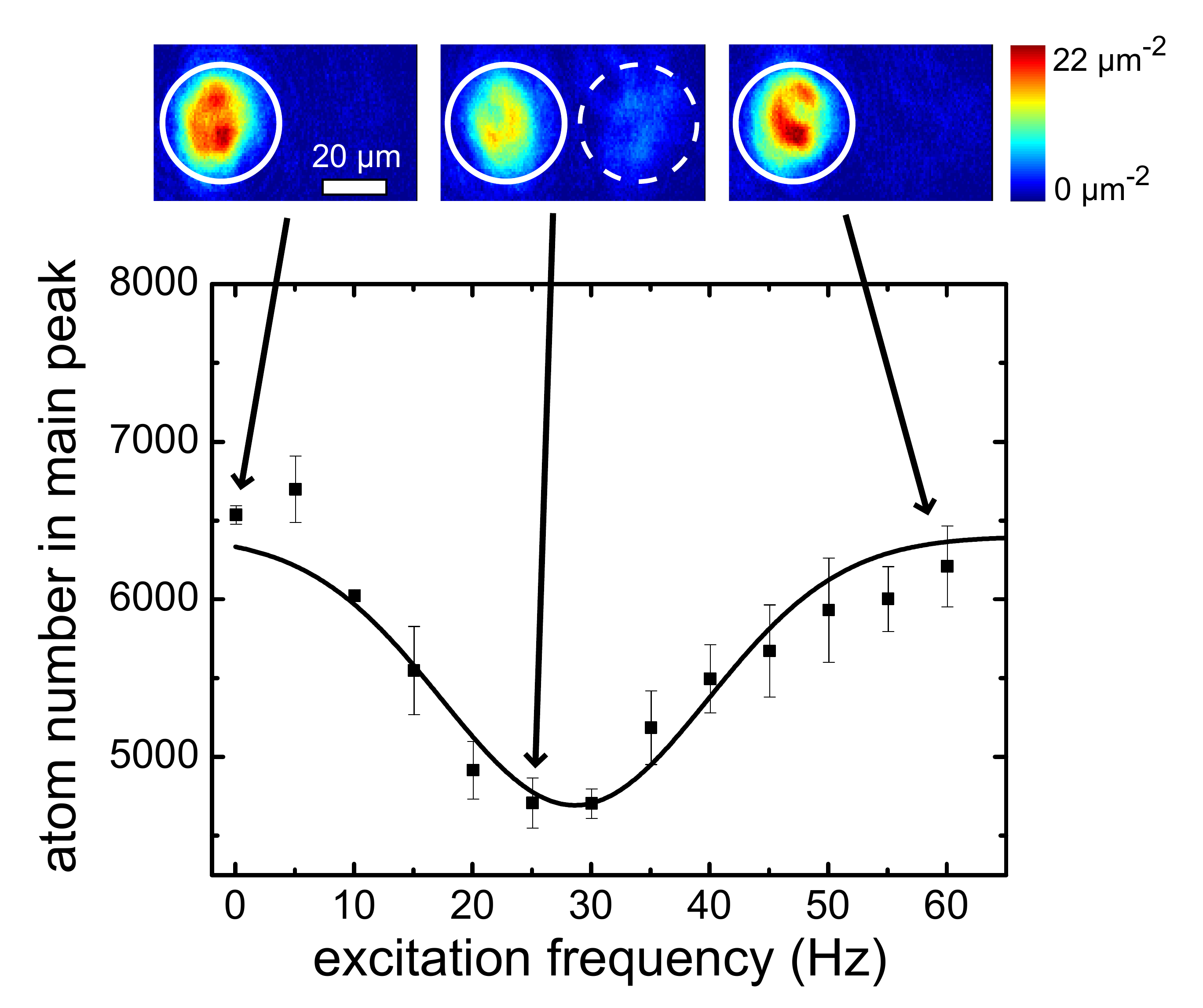}\\
\caption{\textbf{Determining a dispersion point} This plot shows the atom number detected in the main peak after applying an excitation of varying frequency with $k=0.44~k_{\rm{L}}$ to an atomic sample with $a=13~a_0$. Example images (each the average of 4 or 5 experimental trials) illustrate the TOF results. Diffraction peaks from the lattice are outside of the field of view. The atom number is determined by integrating the signal present in the solid white circle. The central image, corresponding to a near-resonant frequency, has a clearly depleted main peak. The dashed white circle indicates the location where atoms transferred to the roton minimum appear after TOF. The solid curve in the plot is a Gaussian fit which yields the excitation frequency of 29(3) Hz.\label{figS2}}
\end{figure}

The dispersion relation corresponds to the points in wavevector and frequency space at which we observe resonant heating of the atom cloud. We determine those points by probing the atomic sample at a fixed wavevector and scanning the DMD triggering frequency. We typically apply the exciting optical potential to the cloud for 40 ms, then perform 30 ms time of flight (TOF) to determine the number of atoms remaining in the condensate. When the excitation is resonant, atoms are excited out of the condensate, which we can observe as a depletion of the atom number in the momentum state of the condensate after TOF. Fig.~\ref{figS2} shows an example loss curve. The fit is to a Gaussian whose center we take to be the resonance frequency. The resonance frequency is not sensitive to the particular fit function chosen: fitting to a Lorentzian instead of a Gaussian typically shifts resonances by tenths of Hz.

The example images in Fig.~\ref{figS2} show the difference between the full and depleted clouds. For this excitation, which at $k=0.44~k_{\rm{L}}$ is near the roton momentum, the atoms missing from the main condensate peak (inside the solid white circles) appear at the right side of the image in a location corresponding to the roton momentum after TOF (inside the dashed white circle). For applied potentials with wavevectors far from the roton minimum, the excited atoms do not always appear in a predictable place. However, the depletion of the condensate remains a consistent signal for all excitation measurements and therefore we use it throughout this work.

\begin{figure}[tbh]
\includegraphics[width=3.7 in]{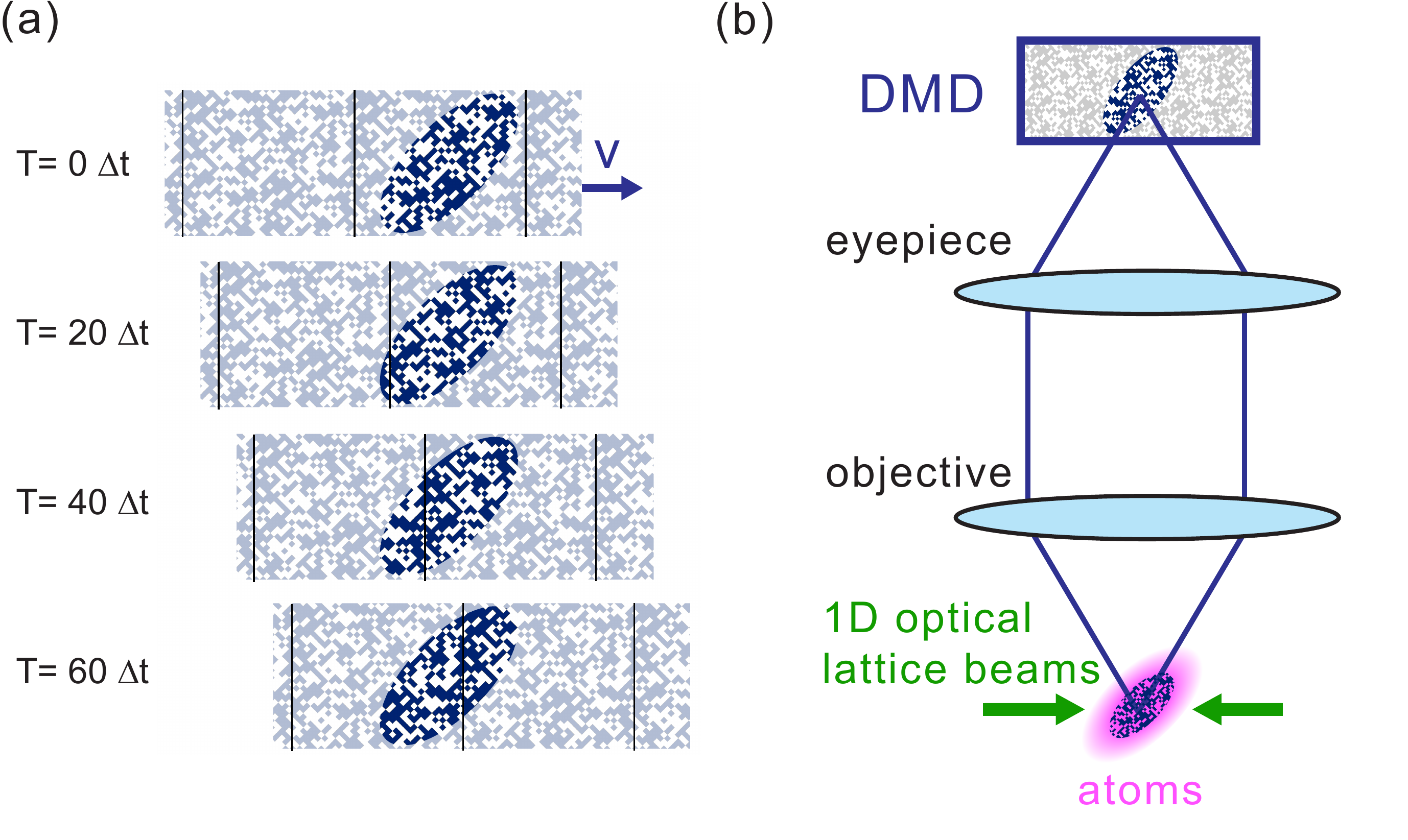}\\
\caption{\textbf{Critical velocity measurement} (a) For critical velocity measurements a digital speckle pattern (light gray) moves at a steady speed $v$ across the atomic sample, but is digitally cropped with the DMD to illuminate only the central region of the sample (dark blue). We show the pattern at four different elapsed times $T=N\times \Delta t$ after the DMD has been triggered $N$ times, where $\Delta t$ is the delay between triggers. The thin vertical lines show the spatial period of the patterns. (b) Critical velocity measurements do not require a beam selection mask. We directly project the real space speckle pattern onto the atomic sample.
\label{figS3}}
\end{figure}

\subsection{Critical velocity measurement}
To measure the critical velocity of the condensate, we move a speckle pattern through the atomic sample as shown in Fig.~\ref{figS3} and determine the minimum velocity required to heat the cloud. Heating is detected using the method shown in Fig.~\ref{figS2}. The speckle pattern is generated on the DMD and directly projected onto the atoms with no beam selection mask in between. However, we do use a digital mask to ensure that we apply the speckle only to the region of high chemical potential, see Fig.~\ref{figS3}(a). We create a speckle pattern by randomly turning on or off $4\times4$ sections of micromirrors instead of individual mirrors. Each section still corresponds to an area smaller than our resolution limit but wastes less laser power into large-angle diffracted peaks that cannot be collected by our projection optics. In principle this speckle pattern should excite the gas at a broad range of wavevectors, limited on the low end by the finite size of the condensate to approximately $k_{\rm{min}}=2\pi/20\mu m \approx 0.05~k_{\rm{L}}$ and limited on the high end by the resolution to approximately $k_{\rm{max}} \approx 0.55~k_{\rm{L}}$.

To move the speckle we trigger the DMD to switch to a different pattern with the same speckle but shifted by $0.3~\mu$m in the plane of the atomic sample. The triggering rate $f=1/\Delta t$ determines the speckle velocity $v=f \times 0.3~\mu$m. Because the DMD can only store up to 96 patterns, the speckle pattern is forced to repeat after $96\times0.3~\mu$m$=29\mu$m which is larger than the width of a sample.

\subsection{Modified Bogoliubov Spectrum}
The shaken lattice is described by a single particle Hamiltonian
\begin{equation}
H_{0}=-\frac{\hbar^{2}}{2m}\frac{d^{2}}{dx^{2}}+V\sin^{2}\left[k_{\rm{L}}\left(x-x_{0}\left(t\right)\right)\right]
\end{equation}
where $m$ is the mass of a particle, $V$ is the lattice depth and $x_{0}\left(t\right)=\left(\Delta x/2\right)\sin\left(\omega t\right)$.
In what follows we will assume an extended system and ignore the harmonic
trapping potential. We use the Trotter expansion to numerically calculate
the single particle spectrum $\epsilon_{0}(q)$ of the time-averaged Hamiltonian \cite{Parker13S}, see Fig.~1(a) in the main text. We will henceforth work in momentum space and project into the single particle band that is adiabatically connected to the $s$-band in the limit of no shaking.

We can describe the interacting Bose gas with the Hamiltonian
\begin{equation}
\hat{H}=\sum_{p}\left[\tilde{\epsilon}_{0}\left(p\right)-\mu\right]\hat{a}_{p}^{\dagger}\hat{a}_{p}+\frac{g}{2v}\sum_{q,p_{1},p_{2}}\hat{a}_{p_{1}+q}^{\dagger}\hat{a}_{p_{2}-q}^{\dagger}\hat{a}_{p_{1}}\hat{a}_{p_{2}},
\end{equation}
where $g$ is the interaction energy, $v$ is the volume of the sample, $\mu$ is the chemical potential and we have applied a gauge transformation to shift the dispersion to $\tilde{\epsilon}_{0}\left(p\right)=\epsilon_{0}\left(p+q^{*}\right)-\epsilon_{0}\left(q^{*}\right)$.
Since the single particle spectrum is asymmetric around the condensate momentum ($p=0$), the standard Bogoliubov formula does not apply. To calculate the excitation spectrum of the system we assume a condensate at $p=0$ and replace the annihilation operator with $\hat{a}_{0}\rightarrow\sqrt{N_{0}}$, where $N_{0}$ is the condensate number. The Bogoliubov Hamiltonian is found by expanding to second order in the fluctuations around the mean-field $\hat{a}_{0}$:
\begin{equation}
\hat{H}_{{\rm Bog}}=\sum_{p\neq0}\left[\left(\tilde{\epsilon}_{0}\left(p\right)+\mu\right)\hat{a}_{p}^{\dagger}\hat{a}_{p}+\frac{\mu}{2}\left(\hat{a}_{p}^{\dagger}\hat{a}_{-p}^{\dagger}+\hat{a}_{p}\hat{a}_{-p}\right)\right],\label{eq:Hbog}
\end{equation}
where $\mu=N_{0}g/v$ and we have neglected an overall mean-field energy shift of the condensate.

To diagonalize $\hat{H}_{{\rm Bog}}$ we define a new set of operators $\hat{b}_{p}$ and $\hat{b}_{-p}^{\dagger}$ implicitly through the relations
\begin{eqnarray}
\hat{a}_{p} & = & u_{p}\hat{b}_{p}+v_{p}\hat{b}_{-p}^{\dagger},\label{eq:Transformation}\\
\hat{a}_{-p}^{\dagger} & = & u_{-p}\hat{b}_{-p}^{\dagger}+v_{-p}\hat{b}_{p},\nonumber
\end{eqnarray}
where we assume $u_{p}$, $v_{p}$ to be real. We require that the Bogoliubov Hamiltonian is diagonal when expressed in terms of the new operators:
\begin{equation}
\hat{H}_{{\rm Bog}}=\sum_{p\neq0}E\left(p\right)\hat{b}_{p}^{\dagger}\hat{b}_{p},
\end{equation}
and that the new operators additionally satisfy the standard commutation
relations: $\left[\hat{b}_{p},\hat{b}_{p^{\prime}}^{\dagger}\right]=\delta_{pp^{\prime}}$,
$\left[\hat{b}_{p},\hat{b}_{p^{\prime}}\right]=0$. We then calculate
the commutators $\left[\hat{a}_{p},\hat{H}_{{\rm Bog}}\right]$,
$\left[\hat{a}_{-p}^{\dagger},\hat{H}_{{\rm Bog}}\right]$, $\left[\hat{b}_{p},\hat{H}_{{\rm Bog}}\right]$,
and $\left[\hat{b}_{-p}^{\dagger},\hat{H}_{{\rm Bog}}\right]$. Imposing the the definition of Eq. (\ref{eq:Transformation}), as well as the constraint that commutation relations are preserved, results in a generalized eigenvalue equation \cite{Pethick}
\begin{equation}
\begin{pmatrix}\tilde{\epsilon}_{0}\left(p\right)+\mu & \mu\\
\mu & \tilde{\epsilon}_{0}\left(-p\right)+\mu
\end{pmatrix}\mathbf{u}\left(p\right)=E\left(p\right)\begin{pmatrix}1 & 0\\
0 & -1
\end{pmatrix}\mathbf{u}\left(p\right),
\end{equation}
where $\mathbf{u}\left(p\right)=\left(u_{p},v_{p}\right)^{T}$. Solving the eigenvalue equation gives the Bogoliubov dispersion shown in Eq.~(1) of the main text. The finite momentum of the condensate breaks the symmetry in momentum around $p=0$.

The Bogoliubov transformation coefficients can be found from the generalized eigenvector:
\begin{equation}
\mathbf{u}\left(p\right)=\frac{1}{N\left(p\right)}\begin{pmatrix}f\left(p\right)+\sqrt{f^{2}\left(p\right)-\mu^{2}}\\
-\mu
\end{pmatrix},
\end{equation}
where $N^{2}\left(p\right)=2\sqrt{f^{2}\left(p\right)-\mu^{2}}\left(\sqrt{f^{2}\left(p\right)-\mu^{2}}+f\left(p\right)\right)$
and $f\left(p\right)=\bar{\epsilon}\left(p\right)+\mu$ normalizes the eigenvector such that $\mathbf{u}^{{\rm T}}\left(p\right)\begin{pmatrix}1 & 0\\
0 & -1
\end{pmatrix}\mathbf{u}\left(p\right)=1$.

\subsubsection{Phonon, maxon, and roton excitations}
As discussed in the main text, and shown in Fig.~1-3, the Bogoliubov spectrum has a linear dispersion near $p/q^{*}\ll1$, followed by a maxon at $p=p_{m}$ and a roton at $p=p_{r}$. To calculate the phonon velocity we expand the Bogoliubov spectrum for small $p$. The dispersion near $p=0$ is
\begin{eqnarray}
E\left(p\right) & \approx & \sqrt{\frac{p^{2}}{2m^{*}}\left(\frac{p^{2}}{2m^{*}}+2\mu\right)}\rightarrow\sqrt{\frac{\mu}{m^{*}}}|p|
\end{eqnarray}
with the effective mass
\begin{equation}
m^{*}=\left(\left.\frac{d^{2}\tilde{\epsilon}_{0}\left(p\right)}{dp^{2}}\right|_{p=0}\right)^{-1}\label{eq:RotonMass}
\end{equation}
which implies that the phonon velocity is given by $v_{s}^{2}=\mu/m^{*}$.

Away from $p=0$, and for sufficiently small $\mu\ll\tilde{\epsilon}_{0}\left(p\right)$, the spectrum can be approximated by expanding in $\mu/\bar{\epsilon}\left(p\right)$:
\begin{equation}
E\left(p\right)\approx\tilde{\epsilon}_{0}\left(p\right)+\mu+\mathcal{O}\left(\left[\frac{\mu}{\bar{\epsilon}\left(p\right)}\right]^{2}\right).\label{eq:BogApprox}
\end{equation}
This implies that the roton and maxon occur near the single particle minimum and maximum respectively. The roton and maxon quasi-momenta are therefore well approximated by $p_{r}\approx 2|q^{*}|$ and $p_{m}\approx|q^{*}|$, with their energies
\begin{eqnarray}
\Delta_{r} & \approx &\mu, \\
\Delta_{m} &\approx &\tilde{\epsilon}_{0}\left(-q^{*}\right)+\mu,
\end{eqnarray}
respectively. In both cases the spectrum is linear in the chemical potential, or equivalently in $\left(a/a_{0}\right)^{2/5}$, as was found in Fig.~3(b) of the main text. 

\subsubsection{Critical velocity by the Landau criteria}
The critical velocity for a superfluid can be found by considering a superfluid moving with velocity $v$ in a reference frame $K$, simultaneously moving at a velocity $v$. The energy of the condensate in frame $K$ is $E=E\left(p\right)$. We apply a Galilean transformation to a reference frame $K^{\prime}$ for which the container of the superfluid is at rest. In frame $K^{\prime}$, the energy of
the condensate is $E=E\left(p\right)+vp+\frac{1}{2}mv^{2}$. We see that in the lab frame, the superfluid is only capable of dissipating energy if $E\left(p\right)+vp<0$. Since $E\left(p\right)$ is positive, this implies that $vp$ must be negative with $|vp|\geq E\left(p\right)$. Furthermore, the critical velocity must occur for the first $p$ that satisfies $|v_{c}p|=E\left(p\right)$. These arguments result in the celebrated Landau criteria:

\begin{equation}
v_{c}= \min_{p}\left|\frac{E\left(p\right)}{p}\right|,\label{eq:CritVelocity}
\end{equation}

\noindent where $\left|E\left(p\right)/{p}\right|$ is the phase speed of the excitation. 

For weak shaking the single particle band structure has a single symmetric minimum at momentum $p=q^{*}=0$. Since the spectrum is symmetric in quasimomentum around $q^{*}=0$, the Bogoliubov dispersion has the standard form of
\begin{equation}
E\left(p\right)=\sqrt{\tilde{\epsilon}_0(p)^2 +2\mu\tilde{\epsilon}_0(p) }.
\end{equation}
The critical velocity in the symmetric case is found by minimizing the phase speed over all $p$. Since there is no roton, the critical velocity is set by the speed of sound at small momenta:
\begin{equation}
v_{c0}=\sqrt{\frac{\mu}{m^*}}.
\end{equation}

As the shaking amplitude increases, the dispersion near this minimum becomes increasingly flat as characterized by $(m^*)^{-1}\rightarrow0$. At the critical shaking above which the double well structure emerges, the quadratic part of this dispersion exactly vanishes and the single particle spectrum is quartic at small $p$. This implies that $E\left(p\right)\propto p^{2}$ so $E\left(p\right)/p\rightarrow0$ as $p\rightarrow0$, and therefore the critical velocity must vanish. This dependence explains the dip in the critical velocity near $\Delta x= 12{\rm nm}$ in Fig.~4(b).

Above the critical shaking amplitude the condensate occupies the minimum at $q^*<0$ and the symmetry of the Bogoliubov spectrum is broken. This asymmetry results in the condensate having two distinct critical velocities in the non-roton and roton directions. The critical velocity in the non-roton direction is set by the phonon velocity, because there is no excitation in that direction with smaller phase speed.  This is found by minimizing the phase speed only for negative momenta

\begin{equation}
v_{c-}=v_{s}=\sqrt{\frac{\mu}{m^{*}}}.
\end{equation}

In the roton direction, on the other hand, rotons can have a smaller phase speed than phonons. Once again the critical velocity is found by minimizing the phase speed, but now for only positive momenta. This function is minimized numerically to produce the dashed red line in Fig.~4(b) above the critical shaking value. For a sufficiently small chemical potential, such that the spectrum away from $p=0$ is well approximated by Eq.~(\ref{eq:BogApprox}), we indeed expect the rotons to have a smaller phase speed than the phonons. This implies the critical velocity in the roton direction is well approximated by
\begin{equation}
v_{c+}\approx \left|\frac{\Delta_{r}}{2q^{*}}\right|.
\end{equation}
We therefore see that the limit of small chemical potential, the ratio of the two critical velocities is given by
\begin{equation}
\left|\frac{v_{c+}}{v_{c-}}\right|\approx\frac{\Delta_{r}/2|q^{*}|}{\sqrt{\mu/m^{*}}}\approx\frac{1}{\sqrt{8}}\sqrt{\frac{\mu}{q^{*2}/2m^{*}}},
\end{equation}
where we have used $\Delta_{r}\approx\mu$ and $q^{*2}/2m^{*}\gg\mu$ for experimentally relevant values. Therefore the critical velocity when moving in the roton direction is significantly smaller than the critical velocity in the non-roton direction. This was observed in Fig.~4(b).

\end{document}